\documentclass[%
	reprint,
	superscriptaddress,
	amsmath,amssymb,
	aps,
	prl,
]%
{revtex4-2}

\usepackage[T1]{fontenc}
\usepackage[utf8]{inputenc}
\usepackage[english]{babel}
\usepackage[autostyle=true]{csquotes}

\usepackage[dvipsnames]{xcolor}
\usepackage{graphicx}
\graphicspath{{figures/}}

\usepackage[%
	caption=false,
	subrefformat=simple,
	labelformat=simple,
]{subfig}

\usepackage{mathtools}
\usepackage{bm}
\usepackage{dsfont}
\usepackage{xspace}
\usepackage{xfrac}

\usepackage{mleftright}
\mleftright % fix spacing issues after \left \right commands

\usepackage[%
	separate-uncertainty=false,
	range-phrase=-,
]{siunitx}

\usepackage{hyperref}
\hypersetup{
	pdftitle={Fraunhofer Patterns in Atomic Josephson Junctions},
	pdfauthor={Kevin T. Geier, Giampiero Marchegiani, Vijay Pal Singh, Juan Polo, Luigi Amico},
}

\usepackage[
	capitalise,
]{cleveref}

\usepackage[dvipsnames]{xcolor}
\usepackage{graphicx}
\graphicspath{{figures/}}

\usepackage[%
	caption=false,
	subrefformat=simple,
	labelformat=simple,
]{subfig}

\usepackage{bm}
\usepackage{mathtools}
\usepackage{dsfont}
\usepackage{xspace}
\usepackage{xfrac}

\usepackage{mleftright}
\mleftright

\usepackage[%
	separate-uncertainty=false,
	range-phrase=-,
]{siunitx}

\usepackage[acronym,nomain]{glossaries}
\glsdisablehyper

\newacronym{bec}{BEC}{Bose--Einstein condensate}
\newacronym{gp}{GP}{Gross--Pitaevskii}
\newacronym{gpe}{GPE}{Gross--Pitaevskii equation}
\newacronym{tf}{TF}{Thomas--Fermi}

\newacronym{dmd}{DMD}{digital micromirror device}

\newacronym{dc}{DC}{direct current}
\newacronym{ac}{AC}{alternating current}

\newacronym{rsj}{RSJ}{resistively shunted junction}
\newacronym{sis}{SIS}{superconductor--insulator--superconductor}
\newacronym{sns}{SNS}{superconductor--normal metal--superconductor}

\newacronym[longplural={Josephson vortices}]{jv}{JV}{Josephson vortex}

\newacronym{1d}{1D}{one-dimensional}
\newacronym{2d}{2D}{two-dimensional}
\newacronym{3d}{3D}{three-dimensional}

\newacronym{sm}{SM}{Supplemental Material}

\DeclareSIUnit\bohr{\text {\ensuremath {a}}_{0}}

\newcommand*{\isotope}[2]{\ensuremath{^{#2}\mathrm{#1}}\xspace}

\newcommand{\e}{\mathrm{e}}
\newcommand{\ii}{\mathrm{i}}
\newcommand{\dd}[1]{\mathop{}\!\mathrm{d}#1}

\DeclarePairedDelimiterX{\commutator}[2]{[}{]}{#1,#2}
\DeclarePairedDelimiterX{\anticommutator}[2]{\{}{\}}{#1,#2}

\DeclareMathOperator{\imag}{Im}

\DeclareMathOperator{\sinc}{sinc}

\DeclarePairedDelimiter{\abs}{\lvert}{\rvert}

\DeclarePairedDelimiterX\norm[1]\lVert\rVert{\ifblank{#1}{\:\cdot\:}{#1}}
\DeclarePairedDelimiterXPP\twonorm[1]{}\lVert\rVert{_2}{\ifblank{#1}{\:\cdot\:}{#1}}
\DeclarePairedDelimiterXPP\infinitynorm[1]{}\lVert\rVert{_{\infty}}{\ifblank{#1}{\:\cdot\:}{#1}}

\DeclarePairedDelimiter{\braket}{\langle}{\rangle}
\DeclarePairedDelimiterX{\ketbra}[2]{|}{|}{#1\delimsize\rangle\delimsize\langle#2}
\DeclarePairedDelimiterX{\overlap}[2]{\langle}{\rangle}{#1\delimsize\vert\mathopen{}#2}
\DeclarePairedDelimiterX{\matrixelement}[3]{\langle}{\rangle}{#1\,\delimsize\vert\,\mathopen{}#2\,\delimsize\vert\,\mathopen{}#3}

\DeclarePairedDelimiterX{\set}[1]{\{}{\}}{%
	
	#1
}

\renewcommand{\vec}{\bm}

\glsunset{1d}
\glsunset{2d}
\glsunset{3d}

\begin{document}

\title{Fraunhofer Patterns in Atomic Josephson Junctions}

\author{Kevin T. Geier}
\email[]{kevin.geier@tii.ae}
\affiliation{Quantum Research Center, Technology Innovation Institute, P.O. Box 9639, Abu Dhabi, United Arab Emirates}

\author{Giampiero Marchegiani}
\affiliation{Quantum Research Center, Technology Innovation Institute, P.O. Box 9639, Abu Dhabi, United Arab Emirates}

\author{Vijay Pal Singh}
\affiliation{Quantum Research Center, Technology Innovation Institute, P.O. Box 9639, Abu Dhabi, United Arab Emirates}

\author{Juan Polo}
\affiliation{Quantum Research Center, Technology Innovation Institute, P.O. Box 9639, Abu Dhabi, United Arab Emirates}

\author{Luigi Amico}
\affiliation{Quantum Research Center, Technology Innovation Institute, P.O. Box 9639, Abu Dhabi, United Arab Emirates}
\affiliation{Dipartimento di Fisica e Astronomia, Universit\`a di Catania, Via S. Sofia 64, 95123 Catania, Italy}
\affiliation{INFN-Sezione di Catania, Via S. Sofia 64, 95127 Catania, Italy}

\date{\today}

\begin{abstract} 
    Driven atomic Josephson junctions allow one to monitor phase-coherent dynamics with unprecedented control and flexibility of the system's physical conditions.
    While cold-atom manifestations of the Josephson effect have been extensively studied in a wide variety of settings, atomic Josephson junctions in synthetic electromagnetic fields remain largely unexplored.
    Here, we show that synthetic magnetic fields can induce Fraunhofer-like modulations of the critical current in atomic Josephson junctions.
    Although this effect presents analogies to the Fraunhofer patterns found in superconducting devices, distinctive features emerge due to the neutral nature of the superfluid.
    We investigate the underlying spatial interference mechanisms and elucidate the role of Josephson vortices in the formation of spatially modulated current distributions based on numerical simulations.
	Our results open up new avenues for matter-wave circuits to deepen our understanding of spatial coherence in Josephson junctions, which are fundamental to the development of novel quantum technologies.
\end{abstract}

%\keywords{}

\maketitle

\begin{figure*}[t]
	\subfloat{\label{fig:fraunhofer:a}}%
	\subfloat{\label{fig:fraunhofer:b}}%
	\subfloat{\label{fig:fraunhofer:c}}%
	\includegraphics[width=\textwidth]{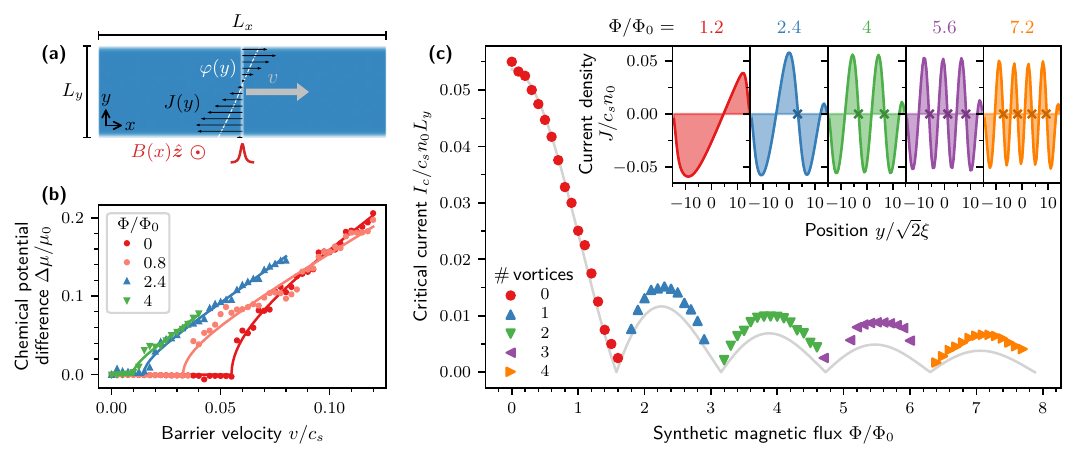}%
	\caption{\label{fig:fraunhofer}%
		Fraunhofer pattern in an atomic Josephson junction. (a)~The junction is created by a repulsive barrier potential, forming a density depletion (light color) in a quasi-\glsentryshort{2d} \glsentrylong{bec} with particle density~$n_0$ and dimensions $L_x \times L_y$ (dark color).
		Moving the barrier at a constant velocity~$v$ induces a net particle current $I \approx - v n_0 L_y$ through the junction.
		A synthetic magnetic field $\vec{B} = B(x - vt) \vec{\hat{z}}$ localized around the junction (red curve) induces a gradient of the phase difference~$\varphi(y)$ across the junction (dashed white line) and therefore a spatial modulation of the current density~$J(y)$ at the junction.
		(b)~Chemical potential difference~$\Delta \mu$ versus barrier velocity~$v$ for different values of the synthetic magnetic flux~$\Phi$.
		A sharp increase of $\Delta \mu$ signals the transition from the \glsentryshort{dc} to the \glsentryshort{ac} regime of the junction at the critical velocity~$v_c$, extracted from fits to a square-root threshold (solid lines).
		(c)~Modulation of the critical current $I_c \approx v_c n_0 L_y$ as a function of $\Phi$, exhibiting a Fraunhofer-like interference pattern.
		The light gray curve marks a fitted $\sinc$ shape, typical of Fraunhofer diffraction patterns of light through a rectangular slit, as a guide to the eye.
		Each lobe is associated with an increment of the number of oscillation periods of the current density~$J(y)$ (see inset) and the number of vortices inside the junction (crosses mark the positions of vortex cores).
	}%
\end{figure*}

Macroscopic quantum coherence emerges when quantum effects persist over spatial regions beyond characteristic microscopic length scales~\cite{10.1143/PTP.69.80,ClarkeNobelLecture2025,DevoretNobelLecture2025,MartinisNobelLecture2025} and can be observed, e.g., in superfluids~\cite{pitaevskiiBoseEinsteinCondensation2016}, superconductors~\cite{tinkham,leggett2008quantum} and, more generally, quantum phases of matter~\cite{leggett1991concept,sachdev2023quantum}.
The ability of controlling quantum coherence in artificially fabricated quantum systems lies at the core of mesoscopic physics and quantum technology~\cite{imry1986physics,dowling2003quantum,acin2018quantum,schleich2016quantum}.
A paradigmatic manifestation of quantum coherence is the Josephson effect, describing the dissipationless flow of electric currents between two superconductors separated by a thin insulating barrier~\cite{ClarkeNobelLecture2025,DevoretNobelLecture2025,MartinisNobelLecture2025,Barone,tinkham}.
Timely after the prediction~\cite{Josephson1962} and experimental verification of the Josephson effect~\cite{PhysRevLett.10.230}, modulations of the junction's critical current by an external magnetic field were reported~\cite{PhysRevLett.11.200}.
The latter are often called Fraunhofer patterns since their shape originates from the same mathematical relation as the Fraunhofer diffration pattern of light through a slit.
Historically, the observation of Fraunhofer patterns was significant as a demonstration of macroscopic phase coherence (Cooper pair transport)~\cite{jaklevic1964quantum,Barone} and for an independent confirmation of the magnetic flux quantum~\cite{PhysRevLett.11.200,jaklevic1964quantum}.
Furthermore, the Fraunhofer effect serves as a key diagnostic tool for the characterization of Josephson junctions in superconducting quantum technologies~\cite{borcsok2019fraunhofer,chen2018finite,yabuki2016supercurrent,Ghatak2018FraunhoferBSTS,Rashidi2024VortexSQI}.

Ultracold atoms stand out among the leading quantum-technology platforms due to their exceptional control of the physical conditions such as particle statistics and interactions~\cite{bloch2012quantum,yago2024atomic}, allowing one to realize versatile coherent quantum circuits within the framework of atomtronics~\cite{amico2022colloquium,Polo_2024}.
Extensive studies have explored Josephson physics in ultracold atoms both theoretically~\cite{Raghavan1999, giovanazziJosephsonEffectsDilute2000,Meier_Zwerger_2001,polo2018damping,polo2019oscillations,Singh_2020,PhysRevA.104.023316,Zaccanti2019,Xhani2020, Xhani2020NJP, Gabriel2023, Jahrling2025} and experimentally~\cite{Cataliotti2001,Oberthaler_2005,Levy2007,Moritz_jj_science,LeBlanc2011,Ji_2022,Roati_dc_jj_scince_2015,Spagnolli2017, Burchianti2018,kwonStronglyCorrelatedSuperfluid2020,Roati_acdc_jj_2021, Gan2025}.
In particular, this platform allows one to control matter-wave densities and junction characteristics to explore transport regimes from weak-link to tunneling limits, as well as to realize setups with spatial geometries ranging from \gls{2d} to \gls{3d}~\cite{Pezze2024stabilizing,singhWeaklinkTunnelingRegime2025}.
Recently, the driven dynamics of atomic Josephson junctions has seen a surge of interest~\cite{singhShapiroStepsDriven2024,delpaceShapiroStepsStronglyinteracting2025,bernhartObservationShapiroSteps2025,SinghVR,SinghJPA,Pradhan2025,Zhu2021}, reinforced by the possibility of observing microscopic dynamics experimentally in real time---a significant challenge in solid-state systems~\cite{van2010colloquium}.
While many studies have focused on \emph{temporal} interference phenomena related to the Josephson effect, the \emph{spatial} interference of Josephson currents underlying Fraunhofer patterns has remained largely unexplored in atomic junctions. 

Here, we propose a setup for realizing Fraunhofer-like patterns in atomic Josephson junctions with synthetic gauge fields.
To this end, we design a protocol in which the Josephson current flows in a \emph{neutral} superfluid subjected to a \emph{synthetic} magnetic field, which can be realized in cold atoms by means of various techniques, including rotation~\cite{andersenQuantizedRotationAtoms2006,cooperRapidlyRotatingAtomic2008,fetterRotatingTrappedBoseEinstein2009}, phase imprinting~\cite{andrelczyk2001optical, andersenQuantizedRotationAtoms2006,kumar2048producing,delpaceImprintingPersistentCurrents2022}, and engineered geometric (Berry) phases~\cite{dalibardColloquiumArtificialGauge2011,goldmanLightinducedGaugeFields2014}.
We demonstrate through numerical simulations based on the \gls{gpe}~\cite{pitaevskiiBoseEinsteinCondensation2016} that the synthetic magnetic field can indeed induce modulations of the critical current of the junction, producing Fraunhofer-like patterns.
Our analysis provides insights into the underlying spatial interference mechanisms as well as signatures of Josephson vortices (also known as fluxons)~\cite{Barone,USTINOV1998315}.
While our results are directly relevant for quantum gases and neutral superfluids, they could also help shedding light on the microscopic physics in superconducting Josephson junctions, which are at the core of various quantum technologies.

\textit{Physical origin of Fraunhofer patterns in Josephson junctions.---}%
In superconducting Josephson junctions, the supercurrent density~$J$ is related to the phase difference~$\varphi$ across the junction via $J = J_c \sin \varphi$, where $J_c$ is the critical current density~\cite{Barone}. Fraunhofer patterns, in their most basic form, arise if a magnetic field produces a linearly varying phase difference along the junction, say, in the $y$-direction, $\varphi(y) = k y + \varphi_0$, where the gradient~$k$ is proportional to the magnetic field and $\varphi_0$ is a phase bias.
The critical current~$I_c$, corresponding to the maximum total current supported by the junction, is then given by the Fourier transform of the critical current density~$J_c(y)$ per unit length, $I_c = | \int \dd{y} \, J_c(y) \e^{\ii k y} |$, in analogy to the far-field pattern of light emerging from Fraunhofer diffraction in optics. In particular, for a rectangular barrier and uniform~$J_c(y)$, the modulations of the critical current as a function of the magnetic flux~$\Phi$ exhibit the characteristic pattern $I_c(\Phi) = I_c(0) | \sinc(\pi \Phi / \Phi_0) |$, where $\sinc(x) = \sin(x) / x$ and $\Phi_0$ is the magnetic flux quantum.
The goal of this work is to design and explore an analog of this scenario with neutral atoms using synthetic gauge fields.

\textit{Theoretical framework and protocol.---}%
We consider an atomic Josephson junction consisting of a thin tunneling barrier which divides a quasi-\gls{2d} \gls{bec} into two superfluid reservoirs, as illustrated in \cref{fig:fraunhofer:a}.
The atoms are confined to a rectangular box of dimensions $L_x \times L_y$.
The barrier is modeled as a repulsive Gaussian potential $V(x) = V_0 \, \e^{-(x - x_0)^2 / 2 \sigma^2}$, where $V_0$ is the barrier strength and $w = 2 \sigma$ determines the barrier thickness, which we choose well within the Josephson tunneling regime~\cite{LeBlanc2011,singhWeaklinkTunnelingRegime2025}, while $x_0$ is the (adjustable) barrier position~\cite{Roati_dc_jj_scince_2015,delpaceShapiroStepsStronglyinteracting2025,bernhartObservationShapiroSteps2025}.
The essential ingredient for observing Fraunhofer-like patterns in this configuration is a synthetic magnetic field~$\vec{B}$ penetrating the region around the junction.
Motivated by the synthetic gauge fields successfully realized in cold-atom experiments~\cite{linSyntheticMagneticFields2009a,dalibardColloquiumArtificialGauge2011,goldmanLightinducedGaugeFields2014}, we choose a synthetic vector potential of the form~$\vec{A}(\vec{r}) = A_y(x) \vec{\hat{y}}$ with $A_y(x) = B_0 \mathinner{(x - x_0)} \mathcal{L}^{1/2}(x)$ and $\mathcal{L}(x) = [1 + (x - x_0)^2 / \ell^2]^{-1}$.
By virtue of $\vec{B} = \vec{\nabla} \times \vec{A}$, this corresponds to the synthetic magnetic field $\vec{B}(\vec{r}) = B(x) \vec{\hat{z}}$ with $B(x) = \mathinner{B_0} \mathcal{L}^{3/2}(x)$.
Here, $B_0$ is the peak value of the magnetic field at the position $x = x_0$, $\vec{\hat{y}}$ and $\vec{\hat{z}}$ denote unit vectors in the respective directions, and $\ell$ plays the role of the magnetic penetration depth (which for our synthetic field is externally imposed, in contrast to bulk superconductors, where the Meissner effect determines the London penetration depth~\cite{Barone}).
We note that our results are qualitatively insensitive to the precise functional form of $B(x)$, as they mainly depend on the total flux piercing the junction~\cite{sm}.

In the weakly interacting regime at zero temperature, the system is well described within \glsentrylong{gp} theory~\cite{pitaevskiiBoseEinsteinCondensation2016} in terms of the energy functional
\begin{equation}
	\label{eq:energyfunctional}
	E = \int \dd \vec{r} \, \left\{ \Psi^* \left[ \frac{(\vec{p} - q \vec{A})^2}{2 m} + V(x) \right] \Psi + \frac{g}{2} \left| \Psi \right|^4 \right\} \,,
\end{equation}
where $\Psi(\vec{r}, t)$ is the condensate wave function, normalized to the total particle number $N = \int \dd{\vec{r}} \, n(\vec{r})$ with $n(\vec{r}) = \abs{\Psi(\vec{r})}^2$ the local density of the atoms, $q$ is a synthetic charge~\footnote{The synthetic vector potential and magnetic field enter all physical quantities exclusively in the form $q \vec{A}$ and $q \vec{B}$, such that the synthetic charge~$q$ can be chosen arbitrarily.},
$m$ is the mass of the atoms, and $g$ is the effective \gls{2d} interaction strength~\cite{hadzibabicTwodimensionalBoseFluids2011a}.
Throughout, we work in natural units defined by the mean particle density $n_0 = N / (L_x L_y)$, the (\glsentrylong{tf}) chemical potential $\mu_0 = g n_0$, the healing length $\xi = \hbar / \sqrt{2 m \mu_0}$, the characteristic time $\tau = \hbar / \mu_0$, the speed of sound $c_s = \sqrt{\mu_0 / m}$, and the synthetic magnetic flux quantum~$\Phi_0 = 2 \pi \hbar / q$, where $\hbar$ is the reduced Planck constant.
The numerical values of the parameters used in our simulations are given in the End Matter.

To probe the critical current as a function of the total magnetic flux $\Phi = \iint B \, \dd{x}\dd{y} \approx 2 B_0 \ell L_y$, we impart a relative velocity between the barrier and the superfluid as follows~\cite{giovanazziJosephsonEffectsDilute2000}.
The system is initialized in the ground state for a given value of $B_0$, obtained by numerically minimizing \cref{eq:energyfunctional}.
Then, we compute the dynamics under the \gls{gpe}~\cite{pitaevskiiBoseEinsteinCondensation2016}, $\ii \hbar \partial_t \Psi = \delta E / \delta \Psi^*$, while moving the barrier (and alongside the vector potential) at a constant velocity~$v$ along~$x$ [see \cref{fig:fraunhofer:a}].
Alternatively, one can also phase-imprint the velocity~$-v$ on the superfluid, $\Psi \to \e^{-\ii m v x / \hbar} \Psi$, while keeping barrier and vector potential fixed~\cite{delpaceImprintingPersistentCurrents2022}.

Below the critical velocity~$v_c$, the system is in the direct-current (\glsentryshort{dc}\glsunset{dc}) or zero-voltage regime~\cite{giovanazziJosephsonEffectsDilute2000}, corresponding to a net total particle current $I \approx -v n_0 L_y$ and a vanishing chemical potential difference~$\Delta \mu = g \braket{n}_R - g\braket{n}_L$.
This $\Delta \mu$ is the neutral-superfluid analog of an electrical voltage in charged superfluids.
Here, $\braket{n}_{L(R)}$ denotes the spatial average over the density on the left (right) side of the barrier within the sonic horizon~\cite{sm}.
In the \gls{dc} regime, the system develops a quasi-stationary supercurrent through the junction, which we probe at a reference time $t_{\mathrm{ref}} = L_x / 2 c_s$ chosen to minimize finite-size effects due to sound waves reflected from the barrier~\cite{sm}.
By contrast, for $\abs{v} > v_c$, the system is in the alternating-current (\glsentryshort{ac}\glsunset{ac}) or voltage regime, characterized by an oscillating current and a nonzero value of $\Delta \mu$~\cite{giovanazziJosephsonEffectsDilute2000}.
The transition between the \gls{dc} and \gls{ac} regime gives access to the critical velocity~$v_c$, which we identify with the critical current $I_c \approx v_c n_0 L_y$.
The latter is extracted, as shown in \cref{fig:fraunhofer:b}, by fitting a voltage--current relation inspired by the \gls{rsj} model~\cite{Barone} to the data, $\Delta \mu(v) \propto \sqrt{v^2 - v_c^2}$~\cite{kwonStronglyCorrelatedSuperfluid2020,Roati_acdc_jj_2021,singhShapiroStepsDriven2024}. In the End Matter, we discuss alternative methods for extracting~$I_c$ based on the current--phase relation and analyze the consistency of both techniques.
Further details on our theoretical framework and protocol can be found in the \gls{sm}~\cite{sm}.

\begin{figure}[t]
	\subfloat{\label{fig:microscopicdynamics:a}}%
	\subfloat{\label{fig:microscopicdynamics:b}}%
	\subfloat{\label{fig:microscopicdynamics:c}}%
	\subfloat{\label{fig:microscopicdynamics:d}}%
	\subfloat{\label{fig:microscopicdynamics:e}}%
	\includegraphics[width=\linewidth]{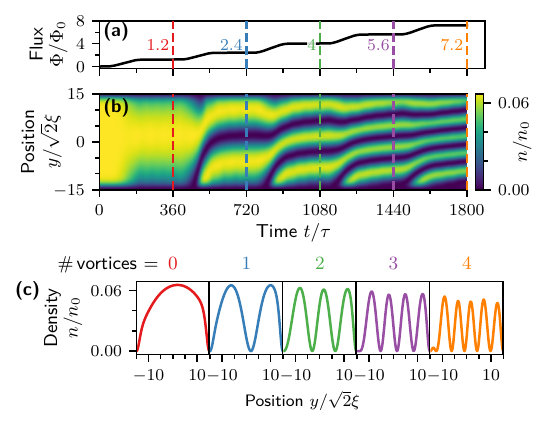}
	\caption{\label{fig:microscopicdynamics}%
		Microscopic dynamics underlying the formation of Fraunhofer patterns in atomic Josephson junctions.
		(a)~The magnetic flux~$\Phi$ is slowly increased in a sequence of consecutive sweeps.
		(b)~Temporal evolution of the density~$n(x_0, y)$ at the junction.
		Each sweep of the magnetic field to a new lobe of the Fraunhofer pattern is accompanied by the entry of an additional vortex into the junction from below, such that the $n$th lobe features $n$ vortices.
		(c)~Snaptshots of the density~$n(x_0, y)$ at values of $\Phi$ marked by vertical dashed lines in panel~(b).
	}
\end{figure}

\textit{Fraunhofer patterns in atomic Josephson junctions.---}%
\Cref{fig:fraunhofer:c} shows the dependence of the critical current on the synthetic magnetic flux, which displays a Fraunhofer-like pattern, as typical for the magnetic-field interference pattern in superconducting Josephson junctions.
This is the main result of our work. An intuitive understanding emerges from the microscopic picture provided by our 
analysis of the underlying current distributions, which explains how the critical current is modulated with increasing magnetic field.
The magnetic field induces a gradient~$k = \partial_y \varphi$ of the phase difference~$\varphi$ across the junction along the $y$ direction, resulting in circulating currents with spatially modulated current density $J(y) \approx J_c(y) \sin(k y + \varphi_0)$ at the junction. The phase bias~$\varphi_0$ is controlled by the barrier velocity and adjusts with increasing $\abs{v} < v_c$ to sustain the externally imposed current. At the critical velocity~$v_c$, the value of $\varphi_0$ is such that the magnitude of the total current $I = \int \dd{y} \, J(y) \approx \imag [ \e^{\ii \varphi_0} \int \dd{y} \, J_c(y) \e^{\ii k y} ]$ through the junction is maximized (typically, $\abs{\varphi_0} = \pi / 2$ for $\abs{v} = v_c$, but see our discussion of dynamical instabilities for our protocol in the \gls{sm}~\cite{sm}). The maximum total current in this configuration is reduced with respect to the maximum current at zero magnetic field due to the spatial modulation of $J(y)$ with wave number~$k$ set by the magnetic field.
As shown in the inset of \cref{fig:fraunhofer:c}, this reduction of the total current can be understood as a spatial interference of left- and right-flowing currents.
If $k$ is such that an integer number of oscillation periods of the current density fit into the junction, this spatial interference is destructive and the critical current vanishes: the system does not support any quasi-stationary supercurrent, but an arbitrarily small bias produces a chemical potential difference and the system enters the \gls{ac} regime.
Conversely, (partial) constructive interference occurs at the maxima of each lobe, fitting a half-integer number of oscillation periods of the current density.

Having outlined the basic mechanisms leading to Fraunhofer-like patterns, it is worthwhile pointing out sources of deviations from the ideal Fraunhofer effect in our setup.
Ideally, the zeros of the Fraunhofer pattern occur at multiples of the magnetic flux quantum~$\Phi_0$, as typically found in rectangular superconducting junctions.
In our case of a neutral superfluid, however, the superflow profile induced by the synthetic magnetic field leads to a reduced gradient of the phase difference~$k < (2 \pi / L_y) (\Phi / \Phi_0)$~\cite{sm}, such that the first zero of the critical current in \cref{fig:fraunhofer:c} occurs at $\Phi > \Phi_0$.
Furthermore, we find that each full oscillation period of the current density corresponds to a vortex with a normal core centered at the position where the current density vanishes [see inset of \cref{fig:fraunhofer:c}].
Each lobe of the Fraunhofer pattern is therefore associated with a specific number of vortices pinned at the junction.
The vortex-induced nonlinearities contribute to deviations of the observed Fraunhofer pattern from the ideal behavior.
In what follows, we analyze the nature of these nonlinearities in more detail.

\textit{Formation of vortices pinned at the junction.---}%
To gain a deeper understanding of the microscopic physics underlying the Fraunhofer-like patterns in our setup, we study in \cref{fig:microscopicdynamics} how the vortices associated with the spatially modulated current density form dynamically.
To this end, as shown in \cref{fig:microscopicdynamics:a}, we slowly ramp up the magnetic field and hold it for a certain amount of time, allowing the system to equilibrate, while moving the barrier at a constant velocity (see End Matter for details on the protocol).
As revealed by \cref{fig:microscopicdynamics:b,fig:microscopicdynamics:c}, each sweep of the magnetic field causes an additional vortex to enter the junction.
In our configuration, the magnetic field produces a phase twist resulting in a clockwise current that is confluent with the negative bias current at the lower boundary, which is why in \cref{fig:microscopicdynamics:b} the vortices enter from the bottom of the junction ($y<0$).

\begin{figure}[t]
	\includegraphics[width=\linewidth]{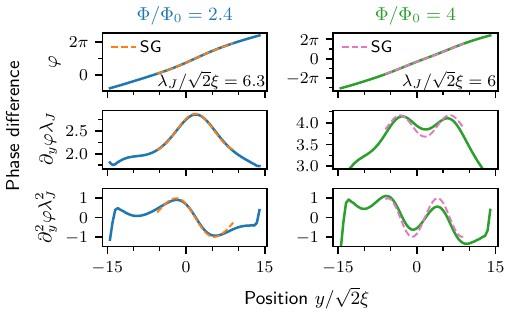}
	\caption{\label{fig:josephsonvortex}%
		Signatures of Josephson vortices.
		The solid lines in each row show the phase difference~$\varphi(y)$ as well as its first and second derivatives for two values of the magnetic flux~$\Phi$ corresponding to the respective configurations in \cref{fig:microscopicdynamics}.
		The location of the phase kinks, visible as peaks in $\partial_y \varphi$, coincides with the position of vortex cores.
		The dashed lines show the right-hand side of the sine-Gordon (SG) equation, $\lambda_J^{-2} \sin \varphi$, with $\lambda_J$ obtained from fits to $\partial_y^2 \varphi$ and integration constants determined by fits to $\partial_y \varphi$ and $\varphi$.
		In the single-vortex case (left column), the kink locally fits the phase profile of a Josephson vortex on a linear background.
	}
\end{figure}

% \textit{Signatures of Josephson vortices.---}%
In superconducting junctions, the appearance of topological defects known as \glspl{jv} or fluxons is a well-known phenomenon~\cite{Barone}.
\Glspl{jv} are localized kinks of the phase difference~$\varphi(y)$, extending transversally across the junction over a length scale set by the Josephson penetration depth~$\lambda_J$.
Mathematically, a \gls{jv} is a solution of the sine-Gordon equation, $\lambda_J^2 \partial_y^2 \varphi = \sin \varphi$, of the form $\varphi_{\mathrm{JV}}(y) = \pm 4 \arctan \exp [(y - y_0) / \lambda_J]$, also known as a sine-Gordon soliton~\cite{USTINOV1998315}, which have also been explored theoretically in atomic condensates~\cite{kaurov20006atomic,su2013kibble,shamailov2018quasiparticles}.

Remarkably, signatures of \glspl{jv} also appear in our implementation of Fraunhofer patterns.
As shown in \cref{fig:josephsonvortex}, these become apparent in the phase difference~$\varphi(y) = (m / \hbar) \int_{x_L}^{x_R} \dd{x} \, v_x(x, y)$ across the junction and its derivatives.
Here, $v_x$ is the $x$-component of the gauge-invariant superfluid velocity~$\vec{v}_s = (\hbar / m) \vec{\nabla} \theta - (q / m) \vec{A}$ with $\theta = \arg(\Psi)$, while the reference points $x_L$ and $x_R$ enclose the density depletion forming the junction (see \gls{sm} for details~\cite{sm}).
Overall, the phase difference exhibits an approximately linear behavior characteristic of Fraunhofer patterns, as described above. 
Deviations from the linear trend consist of localized phase kinks associated with vortex cores, which are more clearly identified as peaks in the derivative~$\partial_y \varphi$.
For the single-vortex case, the profile is consistent with that of a \gls{jv} on a linear background for an effective value of $\lambda_J / \sqrt{2} \xi \approx \num{6}$, which is on the same order of magnitude as the local healing length inside the junction, $\xi^\prime / \xi = \sqrt{n_0 / n_{\mathrm{min}}} \approx \num{4}$.
Deviations from the sine-Gordon model become stronger as the number of vortices increases.

While the emergence of phase kinks and their shape is indicative of \glspl{jv}, the presence of normal cores is unusual compared to the \glspl{jv} found in \glsentryshort{sis} junctions, where there is approximately no suppression of the superconducting order parameter (excluding the insulating barrier)~\cite{Barone,PlastovetsPhysRevB105}.
We remark that the core sizes of the pinned vortices in our configuration is substantially larger than~$\xi$, a feature which distinguishes them from typical vortices appearing in the bulk.
Moreover, we note that indications of \glspl{jv} with normal cores have been found in \glsentryshort{sns}  junctions~\cite{Roditchev2015,chenCurrentInducedHidden2024}.

\textit{Conclusion.---}%
In this work, we have proposed a framework for realizing Fraunhofer patterns in atomic Josephson junctions pierced by a synthetic magnetic field.
We have demonstrated the dynamics of vortices underlying the formation of spatially modulated current distributions, giving insights into microscopic dynamics {that is hard to be accessed} in driven superconducting junctions.
The scheme is agnostic to the implementation of the synthetic gauge field and can be realized by means of established experimental techniques.
While our numerical simulations have been conducted at zero temperature, we expect the qualitative features to persist also at finite temperatures, with possibly different magnitudes of the critical current~\cite{Singh_2020}.

The appearance of Fraunhofer-like patterns in our setup confirms the existence of a spatially phase-coherent Josephson coupling---a phenomenon that can be found remarkably robust whether the carriers are Cooper pairs in a solid-state junction or ultracold atoms in a matter-wave circuit.
A key difference concerns the nature of the coupling to the electromagnetic field: while in charged superfluids the  electromagnetic potentials are dynamical variables intrinsically coupled to the matter through screening and the Meissner effect, static synthetic gauge fields, as considered in this work, are not reciprocally affected by neutral matter~\cite{Meissner_like_PRL118}.
Nevertheless, neutral superfluids can respond to synthetic gauge fields by generating phase gradients that effectively resemble screening currents~\cite{sm}.
As a result, the distinction between long- versus short-junction regimes, characterized by the presence versus absence of nonlinear screening effects, respectively, cannot be stated as a straightforward extension of the reasoning applied in superconductivity~\cite{Barone}.
Peculiarly, our findings for neutral superfluids indicate behavior analogous to superconductors in the short-junction regime displaying Fraunhofer-like patterns, but superimposed with nonlinear effects, e.g., the occurence of Josephson-like vortices, typical of the sine-Gordon-like phase behavior in the long-junction regime.
Specifically, our results indicate the formation of Josephson-like vortices pinned at the junction and characterized by an empty core. These findings may contribute to a deeper understanding of the structure of Josephson vortices, whose relation to Abrikosov vortices remains an active subject of research in superconducting systems~\cite{Gurevich1992,Roditchev2015,PlastovetsPhysRevB102,SatoPRL130,chenCurrentInducedHidden2024}.

Leveraging the versatility of cold atoms, the Fraunhofer effect can be studied in unexplored regimes such as across the BEC--BCS crossover~\cite{Zaccanti2019,Roati_dc_jj_scince_2015} or in the presence of non-Abelian synthetic gauge fields~\cite{dalibard2015introductionphysicsartificialgauge}.
In addition, implementing these effects in atomic SQUIDs~\cite{amico2022colloquium} would represent a further step toward practical applications in quantum technology.

%%%%%%%%%%%%%%%%%%
% End of main text
%%%%%%%%%%%%%%%%%%

\appendix

%%%%%%%%%%%%%%%%%
% Acknowledgments
%%%%%%%%%%%%%%%%%

\begin{acknowledgments}
	\textit{Acknowledgments.---}%
	We thank Gianluigi Catelani and Abbas Hirkani for useful discussions.
\end{acknowledgments}

\bibliography{references}

%%%%%%%%%%%%
% End Matter
%%%%%%%%%%%%

% No page break here. The End Matter directly follows the references section.
% \clearpage

\section{End Matter}

\subsection{Numerical parameters}

For our numerical simulations, we choose physical parameters close to the experiment reported in Ref.~\cite{chauveauSuperfluidFractionInteracting2023a}.
The quasi-\gls{2d} geometry results from a strong harmonic confinement that freezes the motion along the $z$-direction into the harmonic-oscillator ground-state~\cite{hadzibabicTwodimensionalBoseFluids2011a}.
The effective \gls{2d} interaction strength is then given by $g = \tilde{g} \mathinner{\hbar^2 / m}$ with $\tilde{g} = \sqrt{8 \pi} \mathinner{a_s / a_z}$, where $a_s$ is the $s$-wave scattering length and $a_z = \sqrt{\hbar / m \omega_z}$ is the harmonic oscillator length with $\omega_z$ the harmonic oscillator (angular) frequency.
Specifically, we consider a \gls{bec} of $N = \num{9.6e4}$ atoms with dimensionless \gls{2d} interaction constant $\tilde{g} = \num{0.15}$.
The strength of the mean-field interaction thus amounts to $\tilde{g} N = \num{14400}$.
Due to scale invariance in two dimensions, this quantity also sets the system size in natural length units via $L_x L_y / 2 \xi^2 = \tilde{g} N$.
Choosing an aspect ratio of $L_x / L_y = \num{16}$, we have $L_x / \sqrt{2} \xi = \num{480}$ and $L_y / \sqrt{2} \xi = \num{30}$.
Furthermore, we choose the barrier parameters as $V_0 / \mu_0 = \num{3}$ and $w / \sqrt{2} \xi = \num{1}$, deep in the tunneling regime, and we fix the characteristic length scale of the magnetic field as $\ell / \sqrt{2} \xi = \num{5}$.

Let us illustrate the above choice of parameters in SI units for a system of \isotope{Rb}{87} atoms.
The value of $\tilde{g} = \num{0.15}$ is obtained for $a_s \approx \SI{5.3}{\nano\meter}$ and $\omega_z / 2 \pi \approx \SI{3.7}{\kilo\hertz}$.
A realistic physical extent of the box trap consistent with the above parameters may be, for example, $L_x = \SI{160}{\micro\meter}$ and $L_y = \SI{10}{\micro\meter}$.
Then, the characteristic scales become $n_0 = \SI{60}{\per\micro\meter\squared}$, $g n_0 / 2 \pi \hbar \approx \SI[round-mode=figures, round-precision=2]{1.047}{\kilo\hertz}$, $\xi \approx \SI[round-mode=figures, round-precision=2]{0.236}{\micro\meter}$, $\tau \approx \SI[round-mode=figures, round-precision=2]{0.152}{\milli\second}$, and $c_s \approx \SI[round-mode=figures, round-precision=2]{2.19}{\milli\meter\per\second}$.

\subsection{Protocol for studying the formation of vortices pinned at the junction}

Below, we provide details on the dynamical protocol employed for the study of microscopic dynamics in \cref{fig:microscopicdynamics}.
The magnetic field strength~$B_0$ (and thus the magnetic flux~$\Phi$) is slowly increased in a sequence of consecutive sweeps, each consisting of a ramping phase over a time $t_{\mathrm{ramp}} / \tau = \num{240}$ and a holding phase for a time $t_{\mathrm{hold}} / \tau = \num{120}$.
The barrier velocity is ramped up alongside the magnetic field to a value $v / c_s = \num{0.0075}$ during the first sweep, then kept constant.
The step values of~$\Phi$ are chosen representative of the $n$th lobe of the Fraunhofer pattern in \cref{fig:fraunhofer:c} with $n \in \set{0, ..., 4}$.
The ramping times have been chosen sufficiently long to suppress the generation of undesired excitations during the evolution.
The holding time allows the vortices to reposition and approach a quasi-stationary configuration at the junction.
This quasi-stationary \gls{dc} state can be identified with the one that develops locally at the junction following the protocol employed in the context of \cref{fig:fraunhofer} for the respective values of the magnetic flux and barrier velocity.
To suppress finite size effects throughout the evolution in \cref{fig:microscopicdynamics}, we have increased the system extension along $x$ to $L_x / \sqrt{2} \xi = \num{1800}$, while increasing~$N$ to keep the mean density~$n_0$ fixed.

\subsection{Extraction of the critical current via the current--phase relation}

\begin{figure}[t]
	\includegraphics[width=\linewidth]{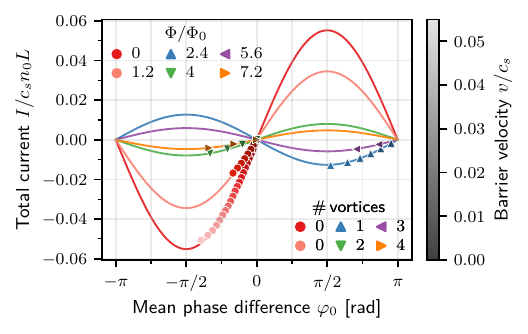}
	\caption{\label{fig:currentphase}%
		Current--phase relation as an alternative method for probing the critical current~$I_c$.
		The solid lines represent fits of the Josephson relation $I(\varphi_0) = I_c \sin \varphi_0$ to the simulation data (markers) with $I_c$ as a fit parameter.
		The barrier velocity~$v$ is encoded in the lightness of the colors.
		For an even (odd) number of vortices inside the junction, the current--phase relation is centered around $\varphi_0 = 0$ ($\varphi_0 = \pi$).%
	}
\end{figure}

\begin{figure}[t]
	\includegraphics[width=\linewidth]{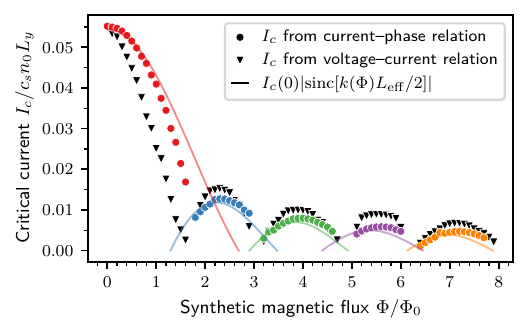}
	\caption{\label{fig:currentphase_vs_voltagecurrent}%
		Comparison of Fraunhofer-like patterns extracted from the current--phase versus voltage--current relation.
		The dots represent the critical current~$I_c$ extracted by fitting the Josephson relation $I(\varphi_0) = I_c \sin \varphi_0$ to the integral current--phase relation in \cref{fig:currentphase}, while the triangles correspond to $I_c \approx \abs{v_c} n_0 L_y$, where the critical velocity~$v_c$ has been extracted from the voltage--current relation as in \cref{fig:fraunhofer:b}.
		In the main lobe, the extrapolated current--phase relation systematically overestimates the critical current realizable in practice with our protocol due to dynamical instabilities.
		The solid lines show the pattern $I_c(\Phi) = I_c(0) \sinc[k(\Phi) L_{\mathrm{eff}} / 2]$, where the wave numbers $k(\Phi)$ have been obtained by fitting the local current--phase relation $J(y) = J_c(y) \sin(k y + \varphi_0)$ to the simulation data (see \gls{sm} for details~\cite{sm}) and deviations from a rectangular critical current density profile~$J_c(y)$ have been accounted for by a fitted effective length~$L_{\mathrm{eff}} / L_y \approx \num[round-mode=figures, round-precision=2]{0.9035}$.
		The color code represents the number of vortices in each lobe, consistent with \cref{fig:fraunhofer:c}.
	}
\end{figure}

Here, we discuss an alternative method for probing the critical current~$I_c$ by exploiting the current--phase relation.
This approach is complementary to the one based on the behavior of the chemical potential difference as a function of the barrier velocity [see \cref{fig:fraunhofer:b}], which we refer to in what follows as the \enquote{voltage--current relation}.
Assuming a Josephson current--phase relation $J(y) = J_c(y) \sin \varphi(y)$ and a linear behavior of the phase difference, $\varphi(y) = k y + \varphi_0$,
the total current through the junction is given by $I(k, \varphi_0) = \imag [ \e^{\ii \varphi_0} \int_{-L_y/2}^{L_y/2} \dd{y} \, J_c(y) \e^{\ii k y} ]$.
If we make the additional assumption that $J_c(y)$ is an even function, which is reasonable for our geometry, this expression simplifies to the familiar Josephson relation for the total current, $I(k, \varphi_0) = I_c(k) \sin \varphi_0$, where the critical current $I_c(k) = \int_{-L_y/2}^{L_y/2} \dd{y} \, J_c(y) \cos(k y)$ is modulated by the magnetic-field-induced gradient~$k$ of the phase difference, resulting in the Fraunhofer interference patterns.

To map out the current--phase relation, we compute the total current $I = \int \dd{y} \, J(y)$ through the junction as well as the mean phase difference $\varphi_0 = L_y^{-1} \int_{-L_y/2}^{L_y/2} \dd{y} \, \varphi(y)$ in a quasi-stationary \gls{dc} state.
As before, such a state is obtained by initializing the system in the ground state and computing the dynamics under the \gls{gpe} while moving barrier and vector potential at a constant velocity up to a reference time chosen such that the relevant observables become approximately stationary~\cite{sm}.
Experimentally, the phase difference~$\varphi_0$ can be measured by inspecting the interference fringes after a time-of-flight expansion during which the two superfluid reservoirs interfere~\cite{Roati_dc_jj_scince_2015,Moritz_jj_science}.
While schemes for the direct or indirect measurement of currents in cold-atom systems exist~\cite{kesslerSinglesiteresolvedMeasurementCurrent2014a,geierNoninvasiveMeasurementCurrents2021a,poloPersistentCurrentsUltracold2025}, the current through the junction can in practice be identified with the imposed barrier velocity via $I \approx -n_0 v L_y$, which is valid deep in the tunneling regime (see \gls{sm} for details~\cite{sm}).

\Cref{fig:currentphase} shows the total current as a function of $\varphi_0$ for several values of the magnetic flux~$\Phi$, representative of specific lobes of the Fraunhofer pattern in \cref{fig:fraunhofer:c}.
The data points include barrier velocities~$v$ sampled on a grid up to the maximum velocity for which the junction is still in the \gls{dc} regime.
For the purposes of this analysis, we consider the system to be in the \gls{dc} regime if the chemical potential difference at the reference time $t_{\mathrm{ref}} = L_x / 2 c_s$ is below a given threshold, $\Delta \mu / \mu_0 \le \num{0.01}$, or if the velocity is less than the critical velocity determined in \cref{fig:fraunhofer:b}.
As can be seen in \cref{fig:currentphase}, the current--phase relation is centered around $\varphi_0 = 0$ ($\varphi_0 = \pi$) for an even (odd) number of vortices inside the junction.
With increasing barrier velocity, $\varphi_0$ decreases with respect to its value at zero current by an amount up to $\pi / 2$, corresponding to the configuration with the maximum supercurrent (in magnitude) supported by the system.
The data is well described by fits to the Josephson relation $I(\varphi_0) = I_c \sin \varphi_0$, with the critical current~$I_c$ as a free parameter for each value of the magnetic flux~$\Phi$.

In \cref{fig:currentphase_vs_voltagecurrent}, we compare the critical current obtained from the current--phase relation, \cref{fig:currentphase}, to the one extracted from the voltage--current relation, \cref{fig:fraunhofer:b}.
It is instructive to analyze the discrepancies between the two methods in more detail.
To this end, consider the current--phase relation in \cref{fig:currentphase} for $\Phi / \Phi_0 = \num{1.2}$, corresponding to a point in the main lobe of the Fraunhofer pattern in \cref{fig:currentphase_vs_voltagecurrent} with a critical current that is significantly reduced with respect to to zero flux.
For this configuration, the data points in \cref{fig:currentphase} do not reach the maximum current extrapolated from the Josephson relation up to $\varphi_0 = \pm \pi / 2$.
The reason for this feature is a dynamical instability: as $v$ increases, the current density locally reaches its critical value, triggering the transition to the \gls{ac} regime even before the bias $\varphi_0$ reaches the configuration that maximizes the magnitude of the supercurrent for a given phase gradient~$k$ (see \gls{sm} for details~\cite{sm}).
In this regime, we can think of the current--phase relation as probing the maximum current that is theoretically supported by the system, while the voltage--current relation probes the maximum current that can practically be reached with our protocol.
We note that this dynamical instability mainly occurs for the lobe with no vortices.
For the other lobes, the agreement between the two evaluation methods improves, while sources of deviations include, for example, fit uncertainties, in particular for the voltage--current method, finite-size effects, inaccuracies in the identification $I \approx -n_0 v L_y$~\cite{sm}, and deviations from the ideal Josephson current--phase relation due to nonlinear effects.
Overall, probing the critical current through the current--phase relation represents an experimentally viable protocol complementary to the voltage--current relation.

\end{document}

\title{Supplemental Material:\texorpdfstring{\\}{} Fraunhofer Patterns in Atomic Josephson Junctions}

\author{Kevin T. Geier}
\email[]{kevin.geier@tii.ae}
\affiliation{Quantum Research Center, Technology Innovation Institute, P.O. Box 9639, Abu Dhabi, United Arab Emirates}

\author{Giampiero Marchegiani}
\affiliation{Quantum Research Center, Technology Innovation Institute, P.O. Box 9639, Abu Dhabi, United Arab Emirates}

\author{Vijay Pal Singh}
\affiliation{Quantum Research Center, Technology Innovation Institute, P.O. Box 9639, Abu Dhabi, United Arab Emirates}

\author{Juan Polo}
\affiliation{Quantum Research Center, Technology Innovation Institute, P.O. Box 9639, Abu Dhabi, United Arab Emirates}

\author{Luigi Amico}
\affiliation{Quantum Research Center, Technology Innovation Institute, P.O. Box 9639, Abu Dhabi, United Arab Emirates}
\affiliation{Dipartimento di Fisica e Astronomia, Universit\`a di Catania, Via S. Sofia 64, 95123 Catania, Italy}
\affiliation{INFN-Sezione di Catania, Via S. Sofia 64, 95127 Catania, Italy}

\date{\today}

\begin{abstract}
	In this Supplemental Material, we provide further information about Fraunhofer patterns in atomic Josephson junctions. In \cref{sec:vectorpotential}, we illustrate the synthetic vector potential employed in our setup, comment on gauge invariance, and discuss the related synthetic electric field generated for a moving barrier.
    \Cref{sec:numericalmethods} puts forward details on our numerical simulations.
    In \cref{sec:observables}, we provide specifics on the definitions and behavior of relevant physical observables, including the current (density), superfluid velocity, phase difference, and chemical potential difference. 
    Finally, in \cref{sec:criticalcurrentdensity,sec:currentdensitymodulation}, we present extended analyses of the local current--phase relation, the critical current density profile, as well as the spatial modulations of the current density, supporting the interpretation of spatial interference as the physical origin of Fraunhofer patterns.
\end{abstract}

%\keywords{}

\maketitle

\setcounter{secnumdepth}{2}

\makeatletter
\renewcommand{\theequation}{S\arabic{equation}}
\renewcommand{\thefigure}{S\arabic{figure}}
\renewcommand{\thetable}{S\arabic{table}}
\renewcommand{\bibnumfmt}[1]{[S#1]}
\renewcommand{\citenumfont}[1]{S#1}
\makeatother

%%%%%%%%%%%%
% SM content
%%%%%%%%%%%%

\section{Synthetic vector potential and gauge invariance}
\label{sec:vectorpotential}

\begin{figure*}[t]
	\subfloat{\label{fig:vectorpotential:a}}%
	\subfloat{\label{fig:vectorpotential:b}}%
	\subfloat{\label{fig:vectorpotential:c}}%
	\subfloat{\label{fig:vectorpotential:d}}%
	\includegraphics[width=\linewidth]{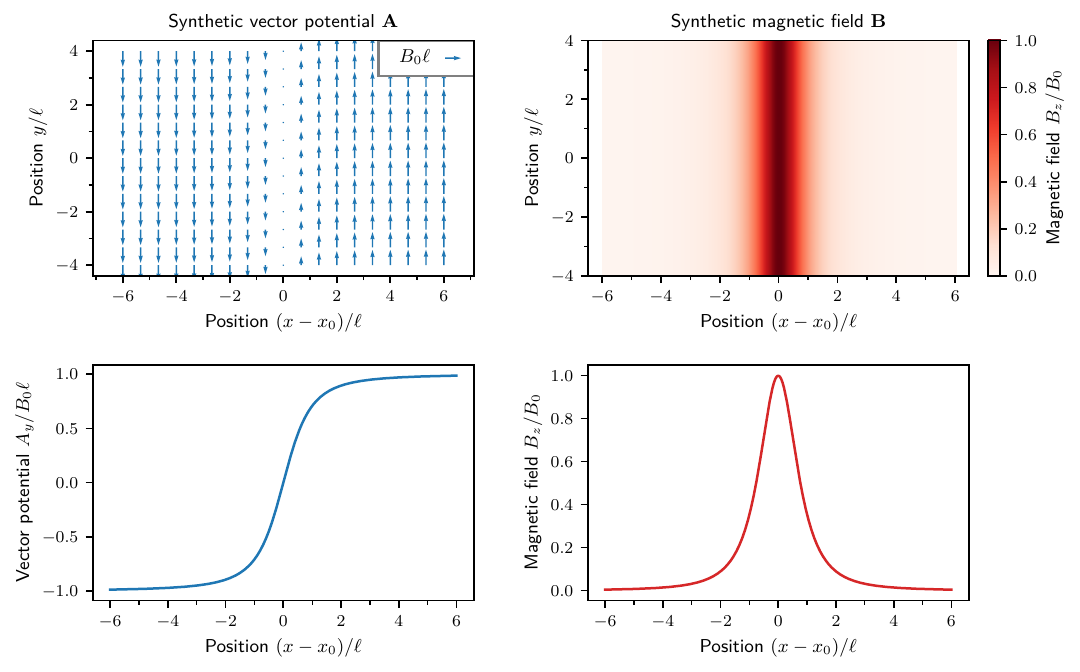}
	\caption{\label{fig:vectorpotential}%
		Synthetic vector potential and synthetic magnetic field according to \cref{eq:synthetic}.%
	}
\end{figure*}

In this section, we provide details on the synthetic vector potential used in our setup.

\subsection{Synthetic vector potential and magnetic field}

We consider the synthetic vector potential
\begin{subequations}
	\label{eq:synthetic}
	\begin{align}
		\label{eq:synthetic:A}
		\vec{A}(\vec{r}) & = B_0 \frac{x - x_0}{\sqrt{1 + (x - x_0)^2 / \ell^2}} \vec{\hat{y}} \,,                                 \\
		\intertext{corresponding to the synthetic magnetic field}
		\label{eq:synthetic:B}
		\vec{B}(\vec{r}) & = \vec{\nabla} \times \vec{A}(\vec{r}) = \frac{B_0}{[1 + (x - x_0)^2 / \ell^2]^{3/2}} \vec{\hat{z}} \,,
	\end{align}
\end{subequations}
as illustrated in \cref{fig:vectorpotential}.
This vector potential is translationally invariant in the $y$-direction and fulfills the Coulomb gauge condition, ${\vec{\nabla} \cdot \vec{A}} = 0$.
Its functional form is motivated by cold-atom experiments that have successfully realized synthetic gauge fields through engineered geometric phases~\cite{linSyntheticMagneticFields2009a,dalibardColloquiumArtificialGauge2011,goldmanLightinducedGaugeFields2014}.
For distances $\abs{x - x_0} \ll \ell$ from the center, the vector potential is approximately linear, $\vec{A} \approx B_0 (x - x_0) \vec{\hat{y}}$, and the magnetic field is approximately uniform, $\vec{B} \approx B_0 \vec{\hat{z}}$, which corresponds to the Landau gauge often considered in quantum Hall physics~\cite{yoshiokaQuantumHallEffect2002a}.

\subsection{Synthetic electric field under motion}

Moving the vector potential alongside the barrier as $\vec{A}(\vec{r}, t) = \vec{A}_0(\vec{r} - vt \vec{\hat{x}} )$ with $\vec{A}_0$ given by \cref{eq:synthetic:A} not only generates the synthetic magnetic field $\vec{B}(\vec{r}, t) = B(x - v t) \vec{\hat{z}}$, but also the synthetic electric field $\vec{E}(\vec{r}, t) = - \partial_t \vec{A}(\vec{r}, t) = v B(x - vt) \vec{\hat{y}}$~\cite{linSyntheticElectricForce2011}.
The system responds to a synthetic electromagnetic force as $\dd{(m \vec{v}_s)} / \dd{t} = q \vec{E} + q \vec{v}_s \times \vec{B}$, where $\vec{v}_s$ is the superfluid velocity.
Thus, our protocol, where barrier and vector potential are dragged along the superfluid at rest, leads in the hydrodynamic limit to a response equivalent to that produced by a Lorentz force $q \vec{E} = q v B \vec{\hat{y}}$, which locally induces a flow of atoms in the $y$-direction.
This is qualitatively consistent with the tilted density profile at the junction found in our simulations, see Fig.~2(c) in the main text.
Note that in the alternative protocol, where barrier and vector potential are at rest and a velocity $\vec{v}_s = -v \vec{\hat{x}}$ is imprinted on the superfluid, there is no synthetic electric field, but the same synthetic Lorentz force $q \vec{v}_s \times \vec{B} = q v B \vec{\hat{y}}$ acts on the atoms.

\subsection{Gauge invariance}

Committing to a specific scheme for realizing synthetic gauge fields with neutral atoms often fixes a specific gauge implicitly.
Nonetheless, the concept of gauge freedom remains of fundamental importance, while physical observables should be gauge invariant. A gauge transformation, as relevant for our scenario, consists of a combined transformation of the vector potential and the wave function according to~\cite{SakuraiNapolitano2017ModernQM}
\begin{subequations}
	\label{eq:gaugetransformation}
	\begin{alignat}{2}
		\label{eq:gaugetransformation:A}
		\vec{A}(\vec{r}, t) & \to \vec{A}^\prime(\vec{r}, t) &  & = \vec{A}(\vec{r}, t) + \vec{\nabla}\chi(\vec{r}, t) \,,   \\
		\label{eq:gaugetransformation:Psi}
		\Psi(\vec{r}, t)  & \to \Psi^\prime(\vec{r}, t)  &  & = \e^{\ii(q/\hbar)\chi(\vec{r}, t)} \Psi(\vec{r}, t) \,,
	\end{alignat}
\end{subequations}
where $\chi(\vec{r}, t)$ is an arbitrary scalar function.
(For completeness, the gauge transformation also involves the synthetic scalar gauge potential, $\phi \to \phi - \partial_t \chi$, which in our case is set to zero.)
It is easy to verify that physical observables, e.g., the density, the superfluid velocity, and derived quantities, are indeed gauge invariant (see \cref{sec:observables}).

\section{Numerical methods}
\label{sec:numericalmethods}

For our purposes, we model the system using \gls{gp} theory, a mean-field approach which is valid if quantum and thermal depletion of the condensate can be neglected.
The dynamics is governed by the \gls{gpe}, which can be derived by applying the variational principle $\ii \hbar \partial_t \Psi = \delta E / \delta \Psi^*$ to the energy functional in Eq.~(1)~\cite{pitaevskiiBoseEinsteinCondensation2016}, yielding
\begin{equation}
	\label{eq:gpe}
	\ii \hbar \partial_t \Psi = \left[ \frac{(-\ii \hbar \vec{\nabla} - q \vec{A})^2}{2 m} + V + g \left| \Psi \right|^2 \right] \Psi \,.
\end{equation}
Here, the vector potential~$\vec{A}$ is incorporated following the minimal coupling prescription.
The ground state~$\Psi_0(\vec{r})$, i.e., the state minimizing the energy functional of Eq.~(1), is a stationary solution of \cref{eq:gpe}, $\Psi(\vec{r}, t) = \Psi_0(\vec{r}) \e^{-\ii \mu t / \hbar}$, satisfying the stationary \gls{gpe}
\begin{equation}
	\label{eq:gpe:stationary}
	\mu \Psi_0 = \left[ \frac{(-\ii \hbar \vec{\nabla} - q \vec{A})^2}{2 m} + V + g \left| \Psi_0 \right|^2 \right] \Psi_0 \,,
\end{equation}
where $\mu$ is the (ground-state) chemical potential.

In our numerical simulations, we solve the \gls{gpe} with homogeneous Dirichlet boundary conditions, i.e., $\Psi(\vec{r}, t) = 0$ on the boundary of the rectangle with corner points $(-L_x/2, -L_y/2)$ and $(L_x/2, L_y/2)$, enforced by a sine pseudospectral discretization~\cite{boydChebyshevFourierSpectral2001,wisePseudospectralTimeDomainPSTD2021}.
Such hard-wall boundaries describe a \gls{bec} trapped in a box potential, which can be generated experimentally with the help of \glspl{dmd}~\cite{villeSoundPropagationUniform2018a,christodoulouObservationFirstSecond2021a}, producing an approximately uniform condensate in the bulk.
For computing the ground state, we use a nonlinear conjugate gradient method~\cite{antoineEfficientSpectralComputation2017a}, while time evolution is implemented via Magnus integration~\cite{blanesMagnusExpansionIts2009}.

\section{Observables}
\label{sec:observables}

\begin{figure*}[t]
	\includegraphics[width=\linewidth]{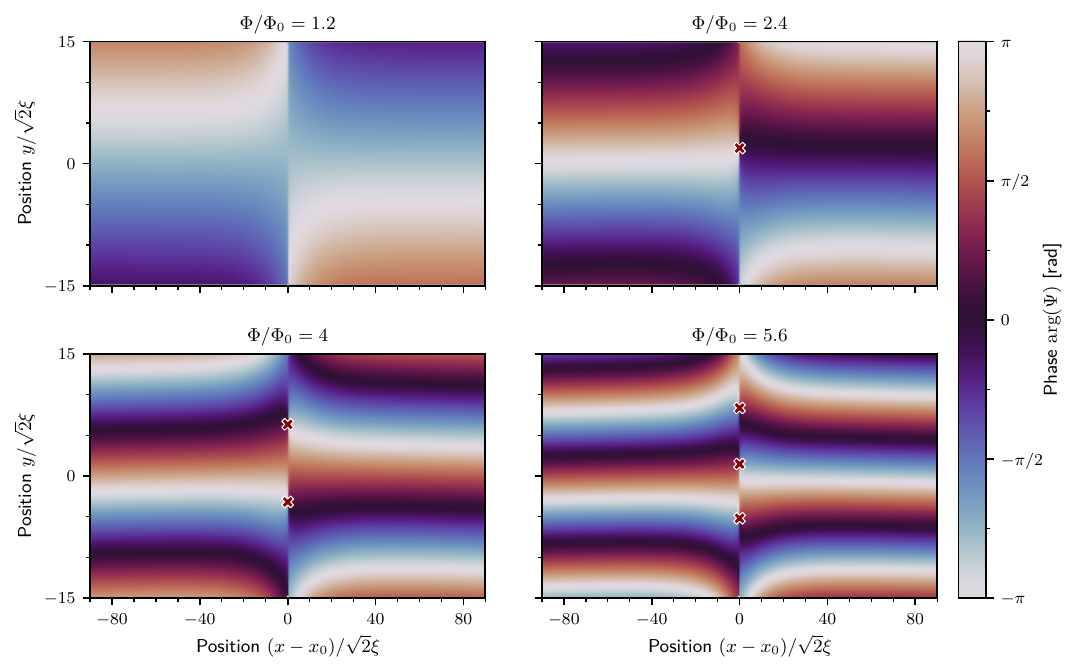}
	\caption{\label{fig:phase}%
		Phase~$\theta = \arg(\Psi)$ of the condensate wave function in a quasi-stationary \glsentryshort{dc} state (barrier velocity~$v / c_s = \num{0.0075}$) for several values of the synthetic magnetic field strength~$B_0$.
		The crosses mark the positions of vortex cores.
	}
\end{figure*}

\begin{figure*}[t]
	\includegraphics[width=\linewidth]{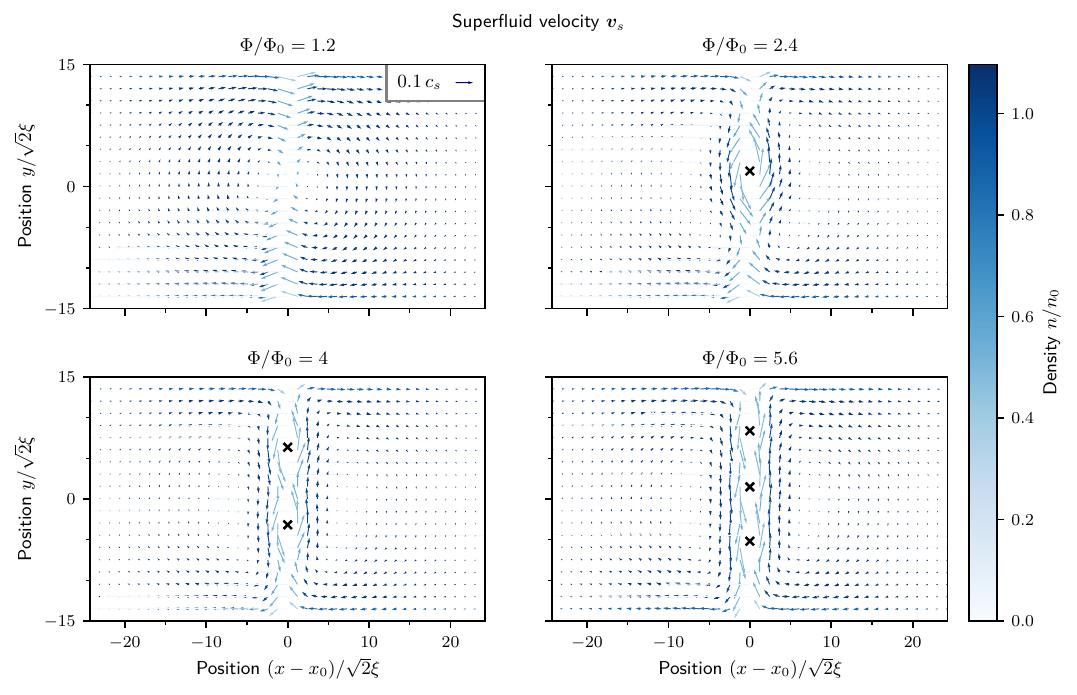}
	\caption{\label{fig:velocity}%
		Superfluid velocity field~$\vec{v}_s$ around the junction in a quasi-stationary \glsentryshort{dc} state (barrier velocity~$v / c_s = \num{0.0075}$) for several values of the synthetic magnetic field strength~$B_0$.
		The crosses mark the positions of vortex cores.
	}
\end{figure*}

\begin{figure}[t]
	\includegraphics[width=\linewidth]{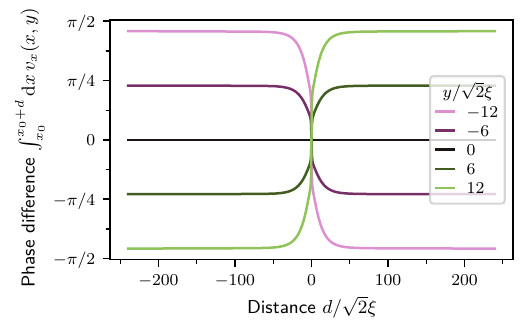}
	\caption{\label{fig:phase_difference_distance}%
		Dependence of the phase difference~$\varphi(y)$ on the distance~$d$ from the position~$x_0$ of the junction along the $x$-direction in the ground state (barrier velocity $v = 0$) for a magnetic flux of $\Phi / \Phi_0 = \num{1.2}$.
	}
\end{figure}

\begin{figure*}[t]
	\includegraphics[width=\linewidth]{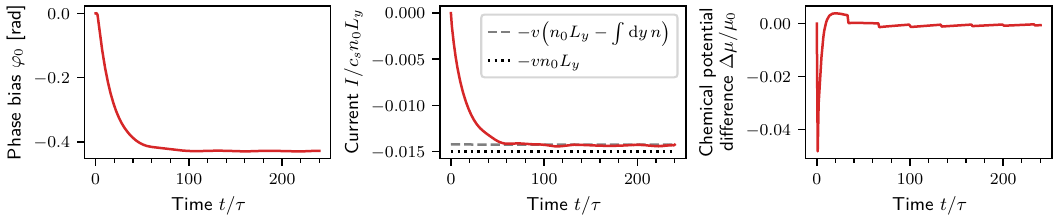}
	\caption{\label{fig:dcrelaxation}%
		Relaxation of the phase bias (mean phase difference)~$\varphi_0$, current~$I$, and chemical potential difference~$\Delta \mu$ in the \glsentryshort{dc} regime for a magnetic flux~$\Phi / \Phi_0 = \num{1.2}$ and barrier velocity~$v / c_s = \num{0.015}$.
		The dashed and dotted horizontal lines in the central panel mark the value of the current estimated from the barrier velocity according to \cref{eq:currentestimate} and the simple relation $I \approx -v n_0 L_y$, respectively. 
	}
\end{figure*}

In this section, we provide details on the definition of observables relevant to our analysis.

\subsection{Current density}

The gauge-invariant current density is defined as
\begin{equation}
	\label{eq:currentdensity}
	\vec{j}(\vec{r}, t) = \frac{1}{m} \real \left[ \Psi^*(\vec{r}, t) \vec{\Pi}(\vec{r}, t) \Psi(\vec{r}, t) \right] \,,
\end{equation}
where $\vec{\Pi}(\vec{r}, t) = \vec{p} - q \vec{A}(\vec{r}, t)$ is the kinetic momentum, formed by the difference between the canonical momentum~$\vec{p} = -\ii \hbar \vec{\nabla}$ and the potential momentum~$q \vec{A}$.
Using the Madelung representation of the wave function, $\Psi(\vec{r}, t) = \sqrt{n(\vec{r}, t)} \e^{\ii \theta(\vec{r}, t)}$, in terms of the density~$n(\vec{r}, t) = \abs{\Psi(\vec{r}, t)}^2$ and the phase~$\theta(\vec{r}, t) = \arg \left[\Psi(\vec{r}, t)\right]$, the current density can be written as
\begin{equation}
	\label{eq:currentdensity:hydro}
	\vec{j}(\vec{r}, t) = n(\vec{r}, t) \vec{v}_s(\vec{r}, t)
\end{equation}
with the gauge-invariant superfluid velocity
\begin{equation}
	\label{eq:superfluidvelocity}
	\vec{v}_s(\vec{r}, t) = \frac{\hbar}{m} \vec{\nabla} \theta(\vec{r}, t) - \frac{q}{m} \vec{A}(\vec{r}, t) \,.
\end{equation}
In what follows, we often omit the temporal argument~$t$ for brevity of notation.

In our discussion of Fraunhofer patterns in atomic Josephson junctions, we typically consider the current density $J(y) = j_x(x_0, y)$ in the hyperplane $x = x_0$, i.e., a vertical line for our two-dimensional geometry, of the junction.
In our protocol this line moves alongside the barrier at a constant velocity~$v$, such that $x_0(t) = vt$.
Note that in our quasi-\gls{2d} geometry, $J(y)$ corresponds to a current density per unit length.
The total (net) current through the junction is then given by $I = \int_{-L_y/2}^{L_y/2} \dd{y} \, J(y)$. In practice, we can obtain the current~$I$ through the junction in the \gls{dc} regime from the externally imposed barrier velocity~$v$ via $I \approx -v n_0 L_y$.
This identification is valid in the deep tunneling regime, where the density inside the junction is low.
We note that a more accurate estimate of the current can be obtained using
\begin{equation}
\label{eq:currentestimate}
	I \approx -v \bigg[ n_0 L_y - \int_{-L_y / 2}^{L_y / 2} \dd{y} \, n(x_0, y) \bigg] \,,
\end{equation}
an expression which takes into account phenomenologically the finite density at the junction.
In principle, this refined expression could be experimentally tested by measuring the density at the junction.
However, for our choice of barrier parameters deep in the tunneling regime, the approximation $I \approx -v n_0 L_y$ is sufficiently accurate in practice for our analysis of Fraunhofer patterns.
In \cref{sec:referenceTime}, we discuss how the current relaxes in the \gls{dc} regime to a quasi-stationary value imposed by the moving barrier.

\subsection{Phase difference}
\label{sec:phasedifference}

To define the phase difference~$\varphi$ across the junction for our quasi-\gls{2d} geometry, we fix reference coordinates $x_L < x_0$ and $x_R > x_0$ on each side of the junction ($x = x_0$ is the hyperplane of the junction).
The gauge-invariant phase difference between any two points $(x_L, y)$ and $(x_R, y)$ with transversal coordinate~$y$ in the hyperplane of the junction is then defined as
\begin{equation}
	\label{eq:phasedifference}
	\begin{split}
		\varphi(y)
		 & = \frac{m}{\hbar} \int_{x_L}^{x_R} \dd{x} \, v_x(x, y) \\
		 & = \theta(x_R, y) - \theta(x_L, y) - \frac{q}{\hbar} \int_{x_L}^{x_R} \dd{x} \, A_x(x, y) \,,
	\end{split}
\end{equation}
where the expression in the second line is obtained by inserting the $x$-component of the gauge-invariant superfluid velocity~$\vec{v}_s$ defined in \cref{eq:superfluidvelocity}.
Note that for our specific choice of gauge, $A_x = 0$, but in general the inclusion of the last term in \cref{eq:phasedifference} is essential for gauge invariance.

A central ingredient of Fraunhofer patterns in Josephson junctions is the magnetic-field-induced spatial variation of the phase difference~$\varphi$ in the hyperplane of the junction, i.e., along the transveral coordinate~$y$.
By taking the $y$-derivative of \cref{eq:phasedifference}, and using $B_z = \partial_x A_y - \partial_y A_x$ together with \cref{eq:superfluidvelocity}, one can derive the expression
\begin{equation}
	\label{eq:phasedifference:gradient}
	\partial_y \varphi(y) = \frac{q}{\hbar} \int_{x_L}^{x_R} \dd{x} \, B_z(x, y) + \frac{m}{\hbar} \left[ v_y(x_R, y) - v_y(x_L, y) \right] \,.
\end{equation}
This result shows that the local variation of the phase difference is determined by the integral of the magnetic field over the interval $[x_L, x_R]$ in the $x$-direction, coarse-graining the small-scale behavior of $B_z$, as well as by contributions from transversal superflows.
As discussed below, the latter are typically negligible in superconducting junctions, while they turn out to be significant in our cold-atom setup.

It is instructive to derive an integral expression for the shift of $\varphi(y)$ along~$y$.
Integration of \cref{eq:phasedifference:gradient} yields
\begin{equation}
	\label{eq:phasedifference:shift:reference}
\begin{split}
	&\varphi(y + \Delta y) - \varphi(y) \\
	&\quad = \hphantom{+} \frac{q}{\hbar} \oint_{\partial R_y(x_L, x_R)} \vec{A}(\vec{r}) \cdot \dd{\vec{r}} \\
	&\quad \hphantom{=} + \frac{m}{\hbar} \int_{y}^{y + \Delta y} \dd{y^\prime} \, \left[ v_y(x_R, y^\prime) - v_y(x_L, y^\prime) \right] \,,
\end{split}
\end{equation}
where the closed line integral in the first term on the right-hand side corresponds to the magnetic flux $\iint_{R_y(x_L, x_R)} B_z \dd{x}\dd{y}$ piercing the rectangle $R_y(x_L, x_R)$ with corner points $(x_L, y)$ and $(x_R, y + \Delta y)$.
This term can be rendered independent of the reference points by extending the closed line integral to the system boundary in the $x$-direction.
After some algebra, one finds
\begin{multline}
	\label{eq:phasedifference:shift}
	\varphi(y + \Delta y) - \varphi(y) \\
	= \frac{q}{\hbar} \oint_{\partial R_y} \vec{A}(\vec{r}) \cdot \dd{\vec{r}} + \frac{m}{\hbar} \oint_{\partial R_y \setminus [x_L, x_R]} \vec{v}_s(\vec{r}) \cdot \dd{\vec{r}} \,.
\end{multline}
Here, the line integral of $\vec{A}(\vec{r})$ runs over the boundary of the rectangle~$R_y$ with corner points $(-L_x / 2, y)$ and $(L_x / 2, y + \Delta y)$, spanning the entire system in the $x$-direction, and corresponds to the magnetic flux $\iint_{R_y} B_z \dd{S}$ through $R_y$.
The line integral of $\vec{v}_s(\vec{r})$ excludes the interval $[x_L, x_R]$ in both horizontal parts of the contour from the integration.

The textbook Fraunhofer effect occurs if the gauge-invariant phase difference varies linearly in the plane of the junction, i.e., $\varphi(y) = k y + \varphi_0$.
In this situation, the gradient~$k$ can be estimated according to \cref{eq:phasedifference:shift} as
\begin{equation}
	\label{eq:phasedifference:gradient:mean}
	k = \frac{\varphi(L_y / 2) - \varphi(-L_y / 2)}{L_y} = \frac{2 \pi}{L_y} \frac{\Phi}{\Phi_0} + \delta k(x_L, x_R) \,,
\end{equation}
where $\Phi = \oint \vec{A} \cdot \dd{\vec{r}} = \iint B_z \dd{x}\dd{y}$ is the total magnetic flux through the system.
The correction $\delta k(x_L, x_R) = L_y^{-1} (m / \hbar) \oint_{\, \Box \setminus [x_L, x_R]} \vec{v}_s \cdot \dd{\vec{r}}$ depends on the reference points and originates from the second term in \cref{eq:phasedifference:shift}, where the integration runs over the system boundary (denoted by $\Box$) excluding the interval $[x_L, x_R]$ in the horizontal parts of the contour.

\subsubsection{Impact of large-scale superflows induced by the synthetic magnetic field}

In the standard treatment of superconducting Josephson junctions, the velocity-dependent terms in \cref{eq:phasedifference:gradient,eq:phasedifference:shift:reference,eq:phasedifference:shift,eq:phasedifference:gradient:mean}, can be made typically negligible~\cite{Barone} .
This is because The Meissner effect expels superflows in the bulk and localizes screening currents near the junction and surfaces, such that transversal components of $\vec{v}_s$ are strongly suppressed at a distance from the junction larger than the London penetration depth.
As a result, the gradient of the gauge-invariant phase difference is given in good approximation by $\partial_y \varphi = (q / \hbar) \int \dd{x} \, B_z$.
The Fraunhofer effect for a rectangular junction in a uniform magnetic field thus produces a modulation of the critical current according to $I_c(\Phi) = I_c(0) \abs{\sinc[k(\Phi) L_y / 2]} = I_c(0) \abs{\sinc(\pi \Phi / \Phi_0)}$ with wave number $k(\Phi) = (2 \pi / L_y) (\Phi / \Phi_0)$.
In particular, the first zero of the critical current occurs for a magnetic flux of $\Phi = \Phi_0$.
We remark that the precise shape of the interference pattern depends also on junction geometry; for instance, in a circular junction, rather than a $\sinc$ function, the critical current follows an Airy pattern with the first node at $\Phi \approx 1.22 \, \Phi_0$.

In our atomic Josephson junction, we find that the phase gradients developed in response to the synthetic magnetic field result to extend well beyond the junction region.
This feature is qualitatively consistent with the expected phase configuration in \gls{2d} neutral superfluids, in which phase distortions propagate logarithmically---see, e.g., Ref.~\cite{nishimori2010elements}.
In charged superfluids, instead, the latter are expected to be suppressed exponentially (on scales of the London penetration depth) due to magnetic screening (see, e.g., Ref.~\cite{tinkham}).
The spatial extension is illustrated in \cref{fig:phase,fig:velocity}, depicting, respectively, the (gauge-variant) phase of the condensate wave function and the (gauge-invariant) superfluid velocity around the junction in a quasi-stationary \gls{dc} state for different values of the synthetic magnetic flux.
Due to these large-scale superflows, the velocity-dependent contributions to \cref{eq:phasedifference:gradient,eq:phasedifference:shift:reference,eq:phasedifference:shift,eq:phasedifference:gradient:mean} are generally not negligible, which can lead to a sizable reduction of the variation of $\varphi(y)$.
In our setup, the spatial modulation of the current density at the junction corresponds to a wave number $k < (2 \pi / L_y) (\Phi / \Phi_0)$ (see discussion in the context of \cref{fig:phase_difference_slope_bias}), such that the first zero of the Fraunhofer pattern occurs for $\Phi > \Phi_0$, see Fig.~1(c).

\subsubsection{Choice of reference points}

If the contributions depending on the tangential velocity components are not negligible in \cref{eq:phasedifference:gradient,eq:phasedifference:shift:reference,eq:phasedifference:shift,eq:phasedifference:gradient:mean}, the phase difference as well as its spatial variation depend on the choice of the reference points $x_L$ and $x_R$.
In \cref{fig:phase_difference_distance}, we show the dependence of the phase difference~$\varphi(y)$ on the distance~$d$ from the junction for $v=0$ and $\Phi=\Phi_0$, corresponding to the choice $x_L = x_0$ and $x_R = x_0 + d$. 
Far away from the junction ($\abs{d} / \sqrt{2} \xi \gtrsim 15$ for our parameter choice), the magnetic-field-induced superflows eventually fade (see \cref{fig:velocity}) and the phase difference in \cref{fig:phase_difference_distance} approaches a constant for each position~$y$.
Thus, for $|x_{L} - x_0|$ and $|x_{R} - x_0|$ sufficiently large, one recovers the reference-independent result~$\partial_y \varphi = (q / \hbar) \int_{-L_x/2}^{L_x/2} \dd{x} \, B_z$ for the gradient of the phase difference, see \cref{eq:phasedifference:gradient}.
However, choosing the reference points in this way would not match the variation of the current density modulations at the junction in our setup, see \cref{sec:currentdensitymodulation}.
Since the boundary of the region across which the phase drop accumulates is not sharp, unlike in SIS junctions, there remains a slight ambiguity in the definition of the local phase difference~$\varphi(y)$ as the choice of reference points is not unique.

For our purposes, we choose $x_L$ and $x_R$ heuristically as the three-sigma interval of a Gaussian profile $n(x) = n_0 - \delta n \, \e^{-(x - x_0)^2 / 2 \sigma^2}$ fitted to the density depletion forming the junction. 
For our parameters, this procedure yields an effective junction width of $\abs{x_R - x_L} / \sqrt{2} \xi \approx \num[round-mode=figures, round-precision=2]{5.996}$ in the absence of a synthetic magnetic field, which slightly increases with increasing magnetic flux up to $\abs{x_R - x_L} / \sqrt{2} \xi \approx \num[round-mode=figures, round-precision=2]{6.516}$ for $\Phi / \Phi_0 = \num{8}$.
As can be seen in \cref{fig:phase_difference_slope_bias}, the gradient of the phase difference computed with this choice of reference points matches approximately the wave number extracted from the modulations of the current density. 

Furthermore, we note that the phase difference on the symmetry line $y = 0$ as well as the phase bias (or mean phase difference) $\varphi_0 = L_y^{-1} \int_{-L_y/2}^{L_y/2} \dd{y} \, \varphi(y)$, which enters the integral Josephson relation $I = I_c \sin \varphi_0$, are largely independent of the choice of reference points: we have checked that $\varphi_0$ is consistent for different reference distances, as long as the interval $[x_L, x_R]$ covers a sufficiently large part of the tunneling region.

\subsection{Chemical potential difference}

The chemical potential difference~$\Delta \mu$ across the junction is our main figure of merit for distinguishing between the \gls{dc} and \gls{ac} regime of the junction, see Fig.~1(b).
To define this quantity, we consider the local chemical potential~$\mu(\vec{r}, t)$, which, within the \glsentrylong{tf} approximation, is given in terms of the local density by $\mu(\vec{r}, t) = g n(\vec{r}, t)$.
In our protocol, we start from the ground state with equal (bulk) densities on both sides of the junction.
In the \gls{ac} regime, the system builds up an imbalance in the density, which propagates from the center of the junction at $x = x_0(t)= v t$ bidirectionally along~$x$ at roughly the speed of sound~$c_s$, while the density outside of this sonic horizon $|x-x_0|=c_s t$ remains at its equilibrium value.
We therefore compute the mean density on the left and right side of the junction as the spatial average within the sonic horizon $\abs{x - x_0(t)} \le c_s t$ according to
\begin{subequations}
\begin{align}
	\braket{n}_L(t) &= \frac{1}{L_y c_s t} \int_{x_0(t) - c_s t}^{x_0(t)} \dd{x} \int_{-L_y/2}^{L_y/2} \dd{y} \, n(x, y, t) \,, \\
	\braket{n}_R(t) &= \frac{1}{L_y c_s t} \int_{x_0(t)}^{x_0(t) + c_s t} \dd{x} \int_{-L_y/2}^{L_y/2}  \dd{y} \, n(x, y, t) \,.
\end{align}
\end{subequations}
The chemical potential difference then amounts to $\Delta \mu(t) = g \braket{n}_R(t) - g \braket{n}_L(t)$.

\subsection{Reference time}
\label{sec:referenceTime}
In \cref{fig:dcrelaxation}, we exemplify the relaxation of key observables, i.e., average phase bias, current and chemical potential difference, to quasi-stationary values in the \gls{dc} regime.
The relaxation time is of the order of a few tens of~$\tau$ for the displayed parameters set.
The steady state persists until finite-size effects due to sound waves reflected from the system boundary set in.
Such phononic excitations are emitted from the center of the junction when the barrier is set in motion and propagate along~$x$ roughly at the speed of sound, reaching the system boundary after a time $L_x / 2 c_s$.
We choose this time as the reference time to probe the \gls{dc} state of the system, although any other choice within the quasi-stationary regime yields the same result.
Choosing $t_{\mathrm{ref}}$ as the maximum time beyond which finite-size effects are expected allows the observables to attain their quasi-stationary values to reasonable accuracy.
For our parameters, the reference time corresponds to $t_{\mathrm{ref}} / \tau = \num{240}$.

\section{Local current--phase relation and extraction of the critical current density}
\label{sec:criticalcurrentdensity}

\begin{figure}[t]
	\subfloat{\label{fig:currentphaselocal:a}}%
	\subfloat{\label{fig:currentphaselocal:b}}%
	\includegraphics[width=\linewidth]{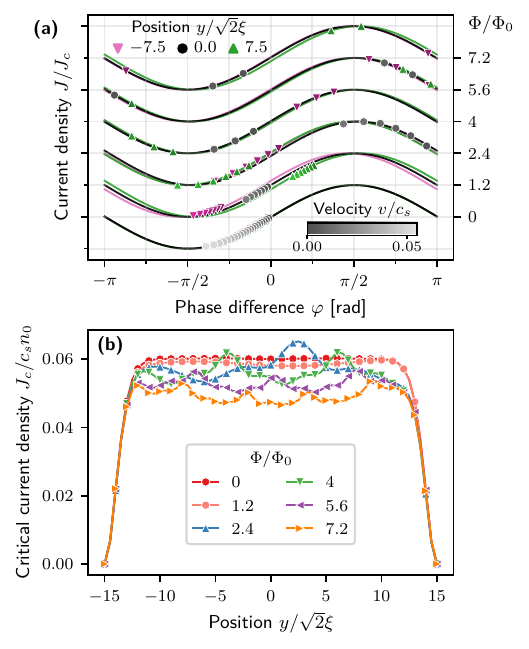}
	\caption{\label{fig:currentphaselocal}%
		Validation of the local Josephson current--phase relation and extraction of the critical current density.
		(a)~The data points show the local current density~$J(y)$ at the junction versus the local phase difference~$\varphi(y)$, computed for a range of barrier velocities~$v$ within the \glsentryshort{dc} regime.
		The data is normalized by the critical current density~$J_c(y)$ and shown for various values of the synthetic magnetic flux~$\Phi$, shifted along the vertical axis.
		Here, $J_c(y)$ is extracted from fits to the model $J(y) = J_c(y) \sin[\varphi(y) + \delta(y)]$ (solid lines).
		Different colors correspond to different positions~$y$, while the lightness encodes the barrier velocity~$v$.
		(b)~Critical current density~$J_c(y)$ corresponding to the fits in panel~(a).
		The spatial profile remains approximately rectangular and does not change significantly with the magnetic field, except for minor features at the locations of the vortex cores.
	}
\end{figure}

The analysis for estimating $I_c$ presented in the End Matter is based on the assumption that the Josephson current--phase relation $J(y) = J_c(y) \sin \varphi(y)$ holds \emph{locally}, i.e., on the level of the current density.
Furthermore, the critical current density~$J_c(y)$ is an intrinsic property of the junction, which should ideally be independent of the synthetic magnetic field.
The latter assumption is, in principle, valid only for sufficiently low values of the synthetic magnetic field.
In \cref{fig:currentphaselocal}, we analyze how well these conditions are fulfilled.
To this end, we have computed the current density at the junction, $J(y)$, for increasing barrier velocities~$v$ within the \gls{dc} regime for a given synthetic magnetic flux~$\Phi$.
At each position~$y$, we have fitted the model $J(y) = J_c(y) \sin[\varphi(y) + \delta(y)]$, where $J(y)$ and $\varphi(y)$ are the \gls{gpe} simulation data, while $J_c(y)$ and $\delta(y)$ are fit parameters.
Here, the phase shift~$\delta(y)$ accounts for the ambiguity of the reference points in our definition of the local phase difference~$\varphi(y)$ (see \cref{sec:phasedifference}) and should be of moderate size for our analysis to be consistent.
The fits are illustrated in \cref{fig:currentphaselocal:a} for the normalized quantity~$J(y) / J_c(y)$, while the extracted spatial profile of the critical current density~$J_c(y)$ is depicted in \cref{fig:currentphaselocal:b}.

As can be seen in \cref{fig:currentphaselocal:a}, the data is well described by the model, while the \textit{ad hoc} inclusion of the phase shift~$\delta(y)$ constitutes only a minor correction (we find $\abs{\delta(y)} \le \num[round-mode=figures, round-precision=1]{0.0677} \, \pi$ for $\abs{y} \le L_y / 3$).
This underlines the validity of the local Josephson current--phase relation for our setup and makes plausible the heuristic choice of the reference points in the definition of the phase difference~$\varphi(y)$.
Moreover, the fact that the critical current density profile in \cref{fig:currentphaselocal:b} changes only marginally (with a deviation of $\SI{15}{\percent}$ at $\Phi/\Phi_0 = \num{7.2}$ compared to the zero-field case) demonstrates that the the specific features of the Josephson coupling established via the barrier potential are robust even in the presence of a sizable magnetic field, and the emergence of Fraunhofer-like patterns is indeed a phase-coherent spatial interference phenomenon.

\section{Analysis of spatial current density modulations}
\label{sec:currentdensitymodulation}

\begin{figure*}[t]
	\includegraphics[width=\linewidth]{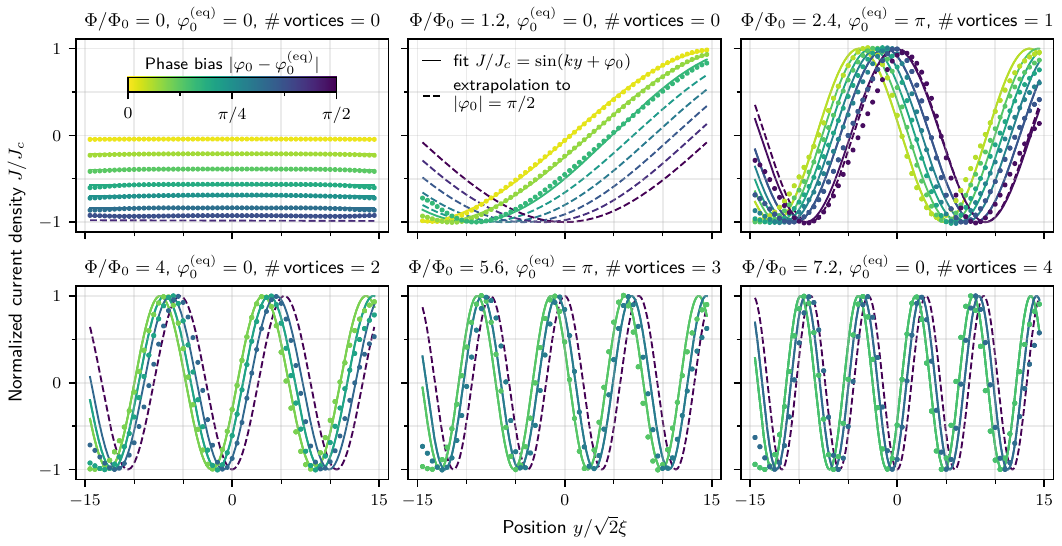}
	\caption{\label{fig:currentdensitymodulation}%
		Spatial modulation of the normalized current density~$J(y) / J_c(y)$ for several values of the synthetic magnetic flux~$\Phi$.
		The local critical current density~$J_c(y)$ has been extracted by fitting the Josephson current--phase relation at each position~$y$ (see \cref{fig:currentphaselocal}).
		The solid lines show the Josephson relation $J(y) / J_c(y) = \sin(k y + \varphi_0)$ with a linear behavior of the phase difference, as relevant for Fraunhofer patterns, where the slope~$k$ and the bias~$\varphi_0$ are fit parameters.
		The data (markers) is shown for increasing values of the barrier velocity up to the critical velocity, corresponding to a change of $\varphi_0$ by up to $\pi / 2$ with respect to its equilibrium value~$\varphi_0^{\mathrm{eq}}$ [$\varphi_0^{\mathrm{eq}} = 0 \, (\pi)$ for an even (odd) number of vortices].
		The dashed lines show extrapolations to the maximum-current configurations at $\abs{\varphi_0} = \pi / 2$, which are not reached within our protocol as the system becomes dynamically unstable with respect to the transition to the \glsentryshort{ac} regime.
	}
\end{figure*}

\begin{figure}[t]
	\subfloat{\label{fig:phase_difference_slope_bias:a}}%
	\subfloat{\label{fig:phase_difference_slope_bias:b}}%
	\includegraphics[width=\linewidth]{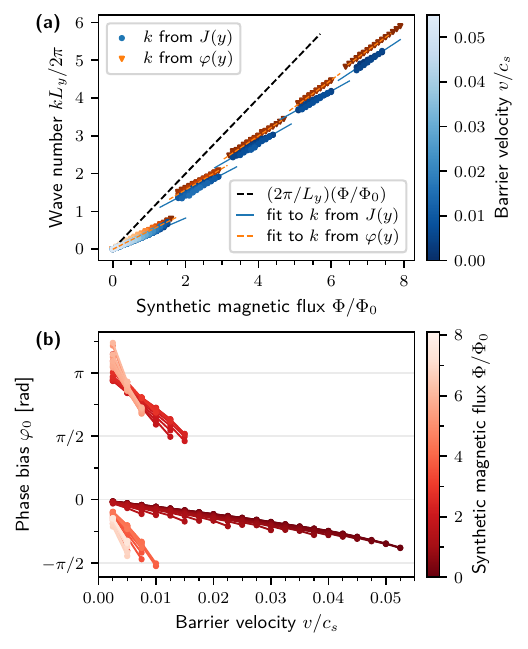}
	\caption{\label{fig:phase_difference_slope_bias}%
		(a)~Wave number~$k$ and (b)~phase bias~$\varphi_0$ characterizing the spatial modulations of the current density at the junction, extracted by fitting the relation $J(y) / J_c(y) = \sin(k y + \varphi_0)$ to the normalized current density (see \cref{fig:currentdensitymodulation}).
		Panel~(a) also includes a comparison of the wave number~$k$ extracted from the current density~$J(y)$ (blue markers) to the corresponding quantity extracted from linear fits to the phase difference~$\varphi(y)$ (orange markers).
		The solid and dashed lines represent linear fits to each cluster, associated with the lobes of the Fraunhofer pattern and the presence of a particular number of vortices.
		The black dashed line shows the relation $k = (2 \pi / L_y) (\Phi / \Phi_0)$, obtained if the reference points are placed far away from the junction (see discussion in \cref{sec:phasedifference}).
	}
\end{figure}

Having established the validity of the local Josephson current--phase relation $J(y) = J_c(y) \sin \varphi(y)$ on the level of the current density, we now go one step further and investigate the magnetic-field-induced modulations of the current density underlying Fraunhofer patterns in more detail.
The Fraunhofer effect occurs if the phase exhibits a linear behavior, $\varphi(y) = k y + \varphi_0$, translating to a spatial modulation of the current density according to $J(y) = J_c(y) \sin(k y + \varphi_0)$.
In \cref{fig:currentdensitymodulation}, we analyze to which extent this relation holds.
To this end, we have conducted sinusoidal fits to the normalized current density, $J(y) / J_c(y) = \sin(k y + \varphi_0)$, where $J(y)$ is obtained from our \gls{gpe} simulation and the critical current density~$J_c(y)$ is extracted following the procedure outlined in \cref{sec:criticalcurrentdensity}, while the wave number~$k$ and the phase bias~$\varphi_0$ are fit parameters.
\Cref{fig:phase_difference_slope_bias} shows the values of $k$ and $\varphi_0$ resulting from these fits.

Overall, for a given value of the magnetic flux~$\Phi$, the fits yield roughly the same value for the wave number~$k$, confirming that the gradient of the phase difference is an approximately linear function of the magnetic field.
The data shown in \cref{fig:currentdensitymodulation} includes barrier velocities~$v$ up to the maximum velocity for which the system is still in the \gls{dc} regime.
Here, we identify the \gls{dc} regime through the criterion $\abs{\Delta \mu(t_{\mathrm{ref}})} / \mu_0 \le 0.01$ or $\abs{v} \le v_c$ with $v_c$ obtained as in Fig.~1(b).
As $v$ increases, the phase bias~$\varphi_0$ changes from its equilibrium value~$\varphi_0^{(\mathrm{eq})}$ at $v = 0$ towards $\abs{\varphi_0} = \pi / 2$, corresponding to the configuration that maximizes the magnitude of the total current $I(k, \varphi_0) = \imag [ \e^{\ii \varphi_0} \int \dd{y} \, J_c(y) \e^{\ii k y} ]$ [assuming a current--phase relation of the form $J(y) = J_c(y) \sin(k y + \varphi_0)$].
Solid lines show fits to valid \gls{dc} configurations, while dashed lines extrapolate to $\abs{\varphi_0} = \pi / 2$ for the same~$k$.
In the absence of a synthetic magnetic field ($\Phi / \Phi_0 = 0$ in \cref{fig:currentdensitymodulation}), the system reaches quasi-stationary \gls{dc} configurations up to values of $\varphi_0$ almost arbitrarily close to $\abs{\varphi_0} = \pi / 2$.
However, for a finite magnetic flux in the main lobe of the Fraunhofer pattern, where vortices are absent ($\Phi / \Phi_0 = \num{1.2}$ in \cref{fig:currentdensitymodulation}), we find that the system transitions to the \gls{ac} regime, as indicated by the $\Delta \mu$ criterion, for values of the phase bias significantly below $\abs{\varphi_0} = \pi / 2$.
We attribute this behavior to a dynamical instability: as $v$ and therefore $\varphi_0$ increases, the system exceeds the critical current density~$J_c(y)$ \emph{locally}, triggering the transition to the \gls{ac} regime before the maximum-current configuration is reached.
This is the physical reason why the critical current extrapolated from the current--phase relation in Fig.~4 can overestimate the value obtained from the voltage--current relation in this regime, see Fig.~5.
Once vortices are present ($\Phi / \Phi_0 \in \set{\num{2.4}, \num{4}, \num{5.6}, \num{7.2}}$ in \cref{fig:currentdensitymodulation}) the stability improves, as indicated by the existence of \gls{dc} configurations with phase biases closer to $\abs{\varphi_0} = \pi / 2$.
While in the absence of vortices the data is well described by the model, deviations from pure sinusoidal behavior can be discerned in the presence of vortices.

If the current density satisfies the relation $J(y) = J_c(y) \sin \varphi(y)$ with $\varphi(y) = k y + \varphi_0$, the wave number characterizing the spatial modulations of the normalized current density $J(y) / J_c(y)$ (see \cref{fig:currentdensitymodulation}) and the gradient of the phase difference~$\varphi(y)$ should coincide.
In \cref{fig:phase_difference_slope_bias:a}, we compare these two approaches for extracting the wave number~$k$.
The data exhibits a piecewise linear behavior with clusters separated by discontinuities that involve a change in offset and slope.
Each cluster is associated with the respective lobe of the Fraunhofer pattern and therefore a specific number of vortices pinned at the junction.
It can be seen that the value of~$k$ extracted from $J(y)$ is close to the one extraced from $\varphi(y)$ in the main lobe of the Fraunhofer pattern (in the absence of vortices).
For higher lobes, deviations become stronger likely due to vortex-induced nonlinearities (see Fig.~3).

Moreover, \cref{fig:phase_difference_slope_bias:a} shows the relation $k = (2 \pi / L_y) (\Phi / \Phi_0)$ (black dashed line), which holds if the reference points $x_L$ and $x_R$ in the definition of the phase difference, \cref{eq:phasedifference}, are chosen far away from the junction where magnetic-field-induced superflows are negligible.
Clearly, such a choice would result in a significant discrepancy to the value of $k$ extracted from the current density modulations, which is independent of the reference points and therefore unambiguous. Instead, by choosing $x_L$ and $x_R$ close to the density depletion forming the junction, as described in \cref{sec:phasedifference}, we find that the gradient of the phase difference approximately agrees with the wave number of the current density modulations.

Summarizing our analysis in the context of \cref{fig:currentphaselocal,fig:currentdensitymodulation,fig:phase_difference_slope_bias}, we can conclude that our atomic Josephson junction in the presence of a synthetic magnetic field approximately follows the local current--phase relation $J(y) = J_c(y) \sin(k y + \varphi_0)$, to the extent discussed in the above.
Since the critical current density profile~$J_c(y)$ is approximately flat (up to  boundary effects), see \cref{fig:currentphaselocal:b}, this suggests that the critical current $I_c = \max_{\varphi_0} \abs{\int \dd{y} \, J(y)}$ should approximately behave as $I_c(\Phi) = I_c(0) \abs{\sinc[k(\Phi) L_{\mathrm{eff}} / 2]}$, where the wave number~$k(\Phi)$ is given by \cref{fig:phase_difference_slope_bias:a} and $L_{\mathrm{eff}}$ is an effective system length that accounts for small deviations from a perfectly uniform critical current density profile.
In Fig.~5, it can be seen that this relation indeed describes the Fraunhofer pattern extracted from the integral current--phase relation in an approximate way.
However, the noticeable deviations hint at physics beyond the textbook Fraunhofer scenario, which becomes manifest, for example, in nonlinear effects due the presence of vortices, as discussed in the main text. 

%%%%%%%%%%%%%%%%%%%
% End of SM content
%%%%%%%%%%%%%%%%%%%

\appendix

\bibliography{references}